\documentclass[twocolumn,amsmath,amssymb,pra]{revtex4}
\usepackage[dvips]{graphicx}
\usepackage{dcolumn}
\usepackage{bm}
\newcounter{one}
\usepackage{braket}
\usepackage[tight]{subfigure}

\begin{document}

\title{Quantum mode filtering of non-Gaussian states\\
for teleportation-based quantum information processing\\}

\author{Shuntaro Takeda}
\email{takeda@alice.t.u-tokyo.ac.jp}
\author{Hugo Benichi}
\author{Takahiro Mizuta}
\author{Noriyuki Lee}
\author{Jun-ichi Yoshikawa}
\author{Akira Furusawa}
\email{akiraf@ap.t.u-tokyo.ac.jp}
\affiliation{Department of Applied Physics, School of Engineering, The University of Tokyo,\\ 7-3-1 Hongo, Bunkyo-ku, Tokyo 113-8656, Japan}

\date{\today}

\begin{abstract}
We propose and demonstrate an effective mode-filtering technique of non-Gaussian states generated by photon-subtraction.
More robust non-Gaussian states have been obtained by removing noisy low frequencies from the original mode spectrum. 
We show that non-Gaussian states preserve their non-classicality after quantum teleportation to a higher degree, when they have been mode-filtered.
This is indicated by a stronger negativity $-0.033\pm0.005$ of the Wigner function at the origin, compared to $-0.018\pm0.007$ for states
that have not been mode-filtered.
This technique can be straightforwardly applied to various kinds of photon-subtraction protocols,
and can be a key ingredient in a variety of applications of non-Gaussian states, especially teleportation-based protocols towards universal quantum information processing. 
\end{abstract}

\maketitle

\section{Introduction}

In the field of continuous-variable (CV) quantum information processing (QIP), 
non-Gaussian states of light have attracted great interest recently as an essential requirement for universal QIP, which is unattainable
only with Gaussian states and operations~\cite{99Lloyd}.
To generate a variety of non-classical non-Gaussian states, the photon-subtraction technique~\cite{97Dakna,06Ourjoumtsev,06Nielsen} has been playing a central role.
With this technique, more advanced and powerful CV protocols with non-Gaussian states have been proposed and also experimentally demonstrated recently~\cite{01Gottesman,11Marek,10Takahashi,11Lee}.

\begin{figure}[!b]
\centering
\subfigure[Sideband modes.]{
\hspace{12mm}\includegraphics[clip,scale=1.0]{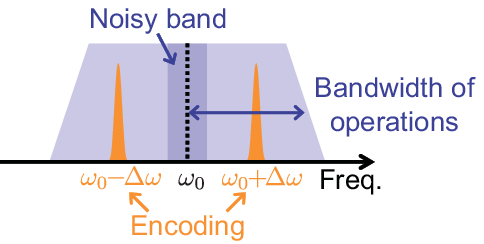}
\label{sfig:sideband_mode}
}
\\
\subfigure[A degenerate mode.]{
\hspace{3mm}\includegraphics[clip,scale=1.0]{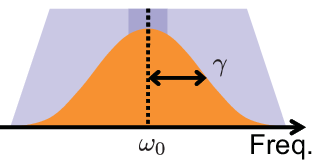}\hspace{3mm}
\label{sfig:degenerate_mode}
}
\subfigure[A filtered degenerate mode.]{
\hspace{5mm}\includegraphics[clip,scale=1.0]{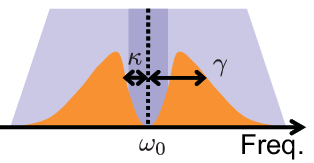}\hspace{5mm}
\label{sfig:filtered_mode}
}
\caption{(Color online) Frequency modes where quantum states are encoded. $\omega_0$:~the carrier frequency, $\gamma$:~the bandwidth of the degenerate mode, $\kappa$:~the cutoff of the optical high-pass-filter.}
\label{fig:mode-filtering}
\end{figure}

In practice, such CV protocols are less efficient in the lower frequencies around the laser carrier frequency
due to various sources of noises, as well as higher frequencies due to the finite bandwidths associated with squeezing,
and also the electronics used.
To achieve the robust implementation of these protocols, fragile non-Gaussian states should be encoded properly within the highly-efficient frequencies. 
However, such encoding has been a challenging task in photon-subtraction, where the frequency mode of non-Gaussian states is determined by a mode of squeezed light associated with probabilistic photon-detection.
Sideband-mode encoding (Fig.~\ref{sfig:sideband_mode}), often used for Gaussian states, can also be achieved here with non-degenerate squeezed modes, but it requires high-speed electronics to observe the generated states~\cite{09Webb}.
In contrast, center-band encoding with a degenerate mode (Fig.~\ref{sfig:degenerate_mode}) is subject to low-frequency noises, which need to be removed with electrical high-pass-filters (HPF) during homodyne-detection.
These HPFs also filter out the low-frequency information of non-Gaussian states,
thereby causing an inevitable frequency mode mismatch which degrades the fragile non-Gaussian states.

Here we propose a mode-filtering strategy to achieve a more efficient encoding of non-Gaussian states via photon-subtraction.
Our strategy is to filter out \textit{optically}
the problematic low-frequencies of the degenerate squeezed mode, instead of using its full spectrum (Fig.~\ref{sfig:filtered_mode}).
By introducing an \textit{optical HPF} with any desired cutoff in the photon-detection channel, we can tailor the frequency mode of the non-Gaussian states without any direct loss on or manipulation of the state itself.
It enables us to achieve almost perfect mode matching in the homodyne-detection, thereby making maximum use of the fragile non-Gaussian states.
In this paper, we develop a theoretical time-domain description of the mode-filtering process, and then experimentally demonstrate the successful mode-reshaping as predicted from the theory. 
The simple technique demonstrated here can be straightforwardly applied to various photon-subtraction protocols, including state-generation~\cite{08Takahashi,07Nielsen} and entanglement distillation~\cite{10Takahashi}.

Our mode-filtering strategy is especially effective for a wide range of CV protocols based on quantum teleportation~\cite{94Vaidman,98Furusawa}, where the low frequencies are often
contaminated by noisy measurements and feed forward operations~\cite{04Yonezawa, 07Yoshikawa, 11Furusawa}.
We have used the quantum teleportation apparatus in Ref.~\cite{11Lee} as a testing device to show the advantages of mode-filtering.
With this device, we teleported non-classical non-Gaussian states generated by mode-filtering, and then measured the degree of non-classicality preserved in the state.
The Wigner function of the teleported state has a minimum negativity of $W(0,0)=-0.033\pm0.005$, compared to $W(0,0)=-0.018\pm0.007$ without mode-filtering, where the negativity here is an indication of the state's non-classicality
($\hbar=1$).
These results show a clear filtering-related enhancement of the quality, thereby demonstrating that our scheme is more effective for the robust implementation of various teleportation-based protocols with non-Gaussian states.

This paper is organized as follows.
In Sec.~\ref{sec:theory}, we first review the basics of photon-subtraction and then introduce the idea and theory of mode-filtering. 
In Sec.~\ref{sec:setup}, we give details of our experimental setup.
In Sec.~\ref{sec:photon-subtraction}, the experimental results of photon-subtraction with mode-filtering are presented and compared with the theory. 
In Sec.~\ref{sec:teleportation}, we show the results of quantum teleportation and demonstrate the advantages of mode-filtering.
Finally Sec.~\ref{sec:conclusion} concludes this paper.



\section{Scheme of mode-filtering}\label{sec:theory}

\begin{figure}[!tb]
\begin{center}
\includegraphics[width=\linewidth]{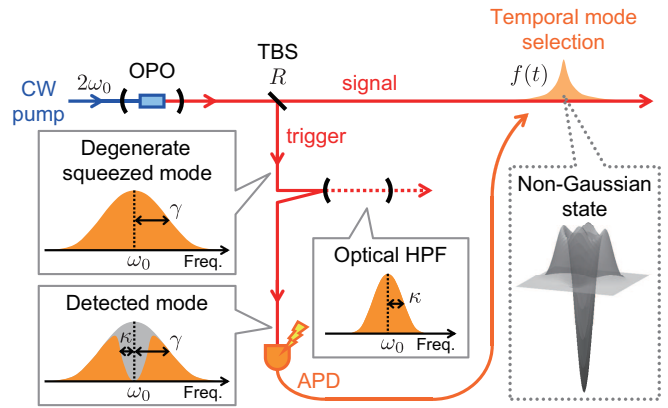}
\end{center}
\vspace{-4mm}
\caption{(Color online) A simple schematic of photon-subtraction with quantum mode-filtering.
APD: Avalanche photo-diode, CW: Continuous wave, HPF: High pass filter, OPO: Optical parametric oscillator, TBS: Tapping beam splitter.}
\label{fig:photon-subtraction}
\end{figure}

Photon-subtraction is a method to conditionally generate non-Gaussian states by the assistance of a single photon detection~\cite{97Dakna}.
As illustrated in Fig.~\ref{fig:photon-subtraction}, a small fraction $R$ of a squeezed vacuum generated by a weakly-pumped OPO is reflected via a tapping beam splitter (TBS) towards an avalanche photo-diode (APD). This beam is used as a trigger beam, whereas the transmitted beam is called the signal beam.
Conditioned on an APD click, one photon is subtracted from its corresponding quantum mode on the signal beam, and the resulting photon-subtracted state shows non-classical features in its non-Gaussian Wigner function with a negative dip.

Since this scheme relies on the quantum correlation between the signal and trigger modes, as well as the projective measurement of the trigger mode,
the frequency mode of the non-Gaussian states is strongly correlated with the frequency mode of the photons detected by the APD.
The degenerate OPO produces correlated photon pairs within a Lorentzian spectrum with bandwidth $\gamma$ in the weak pumping limit.
Any photon within this spectrum can be reflected by the TBS, trigger the APD click, and thereby herald a conditional photon in the signal mode.
Thus, the correlation over the full spectrum is used to induce non-Gaussian states.
As is proven in Ref.~\cite{06Molmer}, the non-Gaussian states have the same frequency spectrum as that of the OPO, and their temporal mode function is given by the Fourier transform of the Lorentzian spectrum as $f_0(t)=\sqrt{\gamma}e^{-\gamma|t|}$ (normalized as $\int\left|f_0(t)\right|^2dt=1$).
In the time domain, this mode is a short wave packet of light, whose information can be extracted by multiplying $f_0(t)$ with the homodyne measurement signal.

In this scheme, the frequency mode of the non-Gaussian state is totally determined by the Lorentzian spectrum of the OPO.
From an engineering point of view, however, this frequency mode can be contaminated by low-frequency laser noise and also electric noise around the carrier frequency $\omega_0$ (more details of the noise contamination are mentioned in Sec.~\ref{sec:setup}).
These types of noise are critical, and strongly deteriorate the fragile non-classicality of the non-Gaussian states.
In order to avoid the noise contamination, we introduce a reflecting cavity before the APD.
The cavity has bandwidth $\kappa$ which transmits photons in the low-frequencies and reflects the rest of the components towards the APD.
The input-output relation of this cavity can be given in the frequency domain by~\cite{00Lu}
\begin{align}
\hat{a}_{\text{r}}(\omega_0+\omega)=\frac{i\omega}{\kappa-i\omega}\hat{a}_{\text{in}}(\omega_0+\omega)
+\frac{\kappa}{\kappa-i\omega}\hat{a}_{\text{v}}(\omega_0+\omega),
\label{eq:cavity_relation}
\end{align}
where $\hat{a}_{\text{in}}$ and $\hat{a}_{\text{r}}$ are annihilation operators of the input and reflected modes, and $\hat{a}_{\text{v}}$ is an auxiliary vacuum mode which does not affect the APD detection.
The response function $i\omega/(\kappa-i\omega)$ in Eq.~(\ref{eq:cavity_relation}) shows that the reflecting cavity acts like an \textit{optical HPF}, which removes low frequency photons before photon-detection.
As a result, quantum correlations in the low frequencies are unaffected by the projective measurement, and thus the non-Gaussian state is generated within a mode without low frequency components.

Note that the optical HPF is introduced not in the signal mode to directly remove the low-frequency components of the state, but in the trigger mode to prevent the low-frequency quantum correlations from being used for inducing non-Gaussian states.
The generation rate of the non-Gaussian state is reduced as a result, but it does not directly affect the target quantum state itself:
as we show below, the ideal heralded state is still the same non-Gaussian state.
Therefore, this scheme makes it possible to reshape the frequency mode of the generated state without any direct loss on or manipulation of the state within the mode.

The mode-filtering scheme can be modeled in time-domain descriptions with an annihilation operator
$\hat{a}(t)=(2\pi)^{-1/2}\int d\omega \hat{a}(\omega_0+\omega)e^{-i\omega t}$.
To begin with, a squeezed state generated from the OPO can be written in the time domain as
\begin{align}
\exp(\hat{P}_{C}-\hat{P}_{C}^\dagger)\ket{0}
=\sum_{n=0}^{\infty}\frac{1}{n!}(\hat{P}_{C}-\hat{P}_{C}^\dagger)^n\ket{0},
\label{eq:squeezedstate}
\end{align}
where
\begin{align}
\hat{P}_{C}^\dagger=\frac{1}{\sqrt{2}}\int dtdt^\prime C(t,t^\prime)\hat{a}^\dagger(t)\hat{a}^\dagger(t^\prime)
\end{align}
is the photon pair creation operator~\cite{00Loudon}.
$C(t,t^\prime)$ denotes the two-time correlation function,
which satisfies $C(t,t^\prime)=C(t^\prime,t)=C(t-t^\prime,0)$ and therefore depends only on the time difference $|t-t^\prime|$ because the output field of the OPO is stationary.
$C(t,t^\prime)$ can be shown to be $O(\epsilon)$, where $\epsilon$ represents the pumping power of the OPO~\cite{00Loudon,84Collett}.

We start with the simplest case of a weak pumping limit of $\epsilon\to0$.
The higher order terms of $n\ge2$ in Eq.~(\ref{eq:squeezedstate}) can be neglected to order $O(\epsilon)$.
The squeezed state thus can be written approximately as $\ket{0}+(\hat{P}_{C}-\hat{P}_{C}^\dagger)\ket{0}=(1-\hat{P}_{C}^\dagger)\ket{0}$, and $\hat{P}_C^\dagger\ket{0}$ is the only term relevant to the photon detection.
By the beam splitter transformation of the TBS $\hat{a}^\dagger(t)\rightarrow \sqrt{1-R}\hat{a}_{\text{s}}^\dagger(t)+\sqrt{R}\hat{a}_{\text{t}}^\dagger(t)$,  $\hat{P}_C^\dagger\ket{0}$ transforms into
\begin{align}
&\frac{1}{\sqrt2}\int dtdt^\prime C(t,t^\prime)\big[(1-R)\hat{a}_{\text{s}}^{\dagger}(t)\hat{a}_{\text{s}}^{\dagger}(t^\prime)\nonumber \\
&+2\sqrt{R(1-R)}\hat{a}_{\text{s}}^\dagger(t)\hat{a}_{\text{t}}^\dagger(t^\prime)
+R\hat{a}_{\text{t}}^{\dagger}(t)\hat{a}_{\text{t}}^{\dagger}(t^\prime)\big]\ket{0}_{\text{s}}\ket{0}_{\text{t}},
\label{eq:tapping}
\end{align}
where the subscripts ``s'' and ``t'' denote the signal and trigger modes, respectively.
Assuming $R\ll 1$, we can neglect the case when more than two photons are reflected towards photon-detection.
Thus, the only term in Eq.~(\ref{eq:tapping}) relevant to the photon-detection has the form
\begin{align}
\int dtdt^\prime C(t,t^\prime)\hat{a}_{\text{s}}^\dagger(t)\hat{a}_{\text{t}}^\dagger(t^\prime)\ket{0}_{\text{s}}\ket{0}_{\text{t}}.
\label{eq:correlated_photons}
\end{align}
Assuming an infinite bandwidth of the APD, we can model an APD detection at a time $t=t_{\text{c}}$ as a projection onto $\bra{0}_{\text{t}}\hat{a}_{\text{t}}(t_{\text{c}})$.
This projective measurement of Eq.~(\ref{eq:correlated_photons}) gives $\int dtC(t-t_{\text{c}},0)\hat{a}_{\text{s}}^\dagger(t)\ket{0}_{\text{s}}$.
The operator $\int dtC(t-t_{\text{c}},0)\hat{a}_{\text{s}}^\dagger(t)$ can be regarded as a creation operator of a single photon, whose temporal mode is defined by a mode function $f(t)\propto C(t-t_{\text{c}},0)$.
In a sense, the beam splitter transformation and the projective measurement subtract one photon from $\hat{P}_C^\dagger\ket{0}$ by replacing the $\hat{P}_C^\dagger$ operator with the corresponding single photon creation operator.
Thus, we can use $f(t)$ to observe the induced single photon state in the weak pumping regime of the conventional photon-subtraction.

When the optical HPF of Eq.~(\ref{eq:cavity_relation}) is applied to the trigger mode before photon-detection,
the effect can be described in the Schr\"{o}dinger picture as the replacement of $\hat{a}_{\text{t}}^\dagger(\omega_0+\omega)\to g(\omega)\hat{a}_{\text{t}}^\dagger(\omega_0+\omega)$, where $g(\omega)=i\omega/(\kappa-i\omega)$.
In Eq.~(\ref{eq:correlated_photons}),
this effect appears as a convolution of $C(t,t^\prime)$ with $g(t)=(2\pi)^{-1/2}\int d\omega g(\omega)e^{-i\omega t}$ as
\begin{align}
\int dtdt^\prime\left(\int d\tau C(t,t^\prime-\tau)g(\tau)\right)\hat{a}_{\text{s}}^\dagger(t)\hat{a}_{\text{t}}^\dagger(t^\prime)
\ket{0}_{\text{s}}\ket{0}_{\text{t}}.
\end{align}
Here the projection onto $\bra{0}_{\text{t}}\hat{a}_{\text{t}}(t_{\text{c}})$ results in
\begin{align}
\int dt \left(\int d\tau C(t-t_{\text{c}}-\tau,0)g(-\tau)\right)\hat{a}_{\text{s}}^\dagger(t)\ket{0}_{\text{s}}.
\end{align}
In this case, the mode function can be defined as
$f_{\text{HPF}}(t)\propto\int d\tau C(t-t_{\text{c}}-\tau,0)g(-\tau)\propto \int d\tau f(t-\tau)g(-\tau)$.
Thus, the effect of the optical HPF can be included as a convolution of $g(-t)$ with the original mode function $f(t)$.

A similar analysis can be applied
to the higher order terms ($n\ge2)$ in Eq.~(\ref{eq:squeezedstate}).
First, a simple inductive proof shows the following statement (A);
``For any order $n$, $(\hat{P}_{C}-\hat{P}^\dagger_{C})^n\ket{0}$ can be decomposed into the terms written only with $n$ or less $\hat{P}^\dagger$ operators with various correlation functions, including the vacuum term $\ket{0}$.''
This statement (A) holds for $n=0$ explicitly. 
Suppose that the statement (A) holds for $n=m$, and therefore $(\hat{P}_{C}-\hat{P}^\dagger_{C})^m\ket{0}$ can be decomposed into  the sum of the terms in the form $\ket{0}$ or $\hat{P}_{C_1}^\dagger\hat{P}_{C_2}^\dagger\cdots\hat{P}_{C_k}^\dagger\ket{0}$ ($k\le m$), where $C_i(t,t^\prime)$ ($i=1,2,\ldots,k$) is the correlation function of each operator.
In this case, all of the terms resulting from the $(m+1)$th order term $(\hat{P}_{C}-\hat{P}^\dagger_{C})^{m+1}\ket{0}$ can be obtained by applying $(\hat{P}_{C}-\hat{P}^\dagger_{C})$ to each $m$th order term as
$(\hat{P}_{C}-\hat{P}^\dagger_{C})\ket{0}=-\hat{P}^\dagger_{C}\ket{0}$ or $(\hat{P}_{C}-\hat{P}^\dagger_{C})\hat{P}_{C_1}^\dagger\hat{P}_{C_2}^\dagger\cdots\hat{P}_{C_k}^\dagger\ket{0}$.
The latter term can be decomposed with the commutation relation $[\hat{a}(t),\hat{a}^\dagger(t^\prime)]=\delta(t-t^\prime)$ and
$[\hat{a}(t),\hat{a}(t^\prime)]=[\hat{a}^\dagger(t),\hat{a}^\dagger(t^\prime)]=0$ as
\begin{align}
&(\hat{P}_{C}-\hat{P}_{C}^\dagger)\hat{P}_{C_1}^\dagger\hat{P}_{C_2}^\dagger\cdots\hat{P}_{C_k}^\dagger\ket{0}\nonumber\\
&=\biggl(\sum_{i}^{k}c_i
\hat{P}_{C_1}^\dagger\cdots\hat{P}^\dagger_{C_{i-1}}
\hat{P}^\dagger_{C_{i+1}}\cdots\hat{P}_{C_k}^\dagger\nonumber\\
&+4\sum_{i<j}^{k}\hat{P}_{C_{i,j}}^\dagger
\hat{P}_{C_1}^\dagger\cdots\hat{P}^\dagger_{C_{i-1}}
\hat{P}^\dagger_{C_{i+1}}\cdots\hat{P}^\dagger_{C_{j-1}}
\hat{P}^\dagger_{C_{j+1}}\cdots\hat{P}_{C_k}^\dagger\nonumber\\
&\hspace{42mm}-\hat{P}_{C}^\dagger\hat{P}_{C_1}^\dagger\cdots\hat{P}_{C_k}^\dagger\biggr)\ket{0},
\label{eq:decomposition}
\end{align}
where $c_i=\int dtdt^\prime C^*(t,t^\prime)C_i(t,t^\prime)$ and
$C_{i,j}(t,t^\prime)=\int d\tau d\tau^\prime C^*(\tau,\tau^\prime)C_i(t,\tau)C_j(\tau^\prime,t^\prime)$. Here, the second term in the right-hand side of Eq.~(\ref{eq:decomposition}) vanishes when $k=1$.
These decompositions show that all of the $(m+1)$th order terms can also be decomposed into the form $\hat{P}_{C_1^\prime}^\dagger\hat{P}_{C_2^\prime}^\dagger\cdots\hat{P}_{C_k^\prime}^\dagger\ket{0}$ ($k\le m+1$), and thus the statement (A) holds for $n=m+1$ as well.
The above discussions prove that the statement (A) holds for any order $n$.

Then, in a similar way, the transformation of the TBS and the projection onto $\bra{0}_{\text{t}}\hat{a}_{\text{t}}(t_\text{c})$ subtract one photon from one of the correlated photon pairs in each term $\hat{P}_{C_1}^\dagger\hat{P}_{C_2}^\dagger\cdots\hat{P}_{C_k}^\dagger\ket{0}$.
As a result, various photon-subtracted terms are produced by the replacement of one of these $\hat{P}^\dagger$ operators with its corresponding single photon creation operator.
The replacement of $\hat{P}_{C_i}^\dagger$ ($i=1,2,\ldots,k$) gives
\begin{align}
&\left(\int dtC_{i}(t-t_{\text{c}},0)\hat{a}_{\text{s}}^\dagger(t)\right)\nonumber\\
&\quad\times\hat{P}_{C_1}^\dagger\hat{P}_{C_2}^\dagger\cdots\hat{P}^\dagger_{C_{i-1}}
\hat{P}^\dagger_{C_{i+1}}\cdots\hat{P}_{C_{k-1}}^\dagger\hat{P}_{C_k}^\dagger\ket{0}_\text{s}.
\label{eq:higher_order_projected}
\end{align}
Here, the projective measurement induces a single photon defined by the mode function $f_i(t)\propto C_{i}(t-t_{\text{c}},0)$.
By selecting the $f_i(t)$ mode, we can observe the pure single photon induced by the APD detection, as well as some fraction of the photon pairs generated in the OPO.
When applying the optical HPF, we need to replace $C_{i}(t-t_{\text{c}},0)$ in Eq.~(\ref{eq:higher_order_projected}) with $\int d\tau C_{i}(t-t_{\text{c}}-\tau,0)g(-\tau)$. The effect of mode-filtering here can also be included as a convolution of $f_i(t)$ with $g(-t)$, as shown previously.
Note that different terms resulting from the projection can have single photons with different mode functions
(when $i\not=j$, $f_i(t)\not=f_j(t)$ in general),
thereby making it impossible to extract all the induced single photons purely at the same time.
However, the mode-filtering effect on all the mode functions $\{f_1(t), f_2(t),\ldots, f_k(t)\}$ can be written in the same way as the convolution with $g(-t)$.
Therefore, when we introduce the optical HPF to the photon-subtraction scheme, we only need to replace the originally-used mode function $f(t)$ with $\int d\tau f(t-\tau)g(-\tau)$, regardless of the pumping power $\epsilon$.

\begin{figure}[!tb]
\begin{center}
\includegraphics[width=0.85\linewidth]{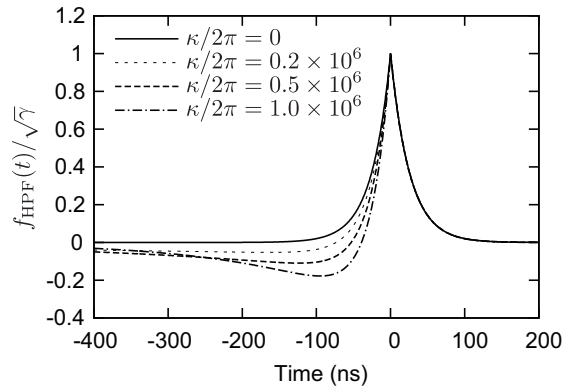}
\end{center}
\vspace{-4mm}
\caption{The dependence of the optically high pass filtered mode function $f_{\text{HPF}}(t)$ on cutoff $\kappa$. We set $\gamma=2\pi\times 6.2$MHz.}
\label{fig:mode_function}
\end{figure}

Our experiment is performed in the weak pumping regime, thus we can assume that the original mode function is $f_0(t)=\sqrt{\gamma}e^{-\gamma|t|}$.
$g(t)$ can be derived from the Fourier transform of $g(\omega)=i\omega/(\kappa-i\omega)$ as
\begin{align}
g(t)&=\sqrt{2\pi}\times
\begin{cases}
\left(\kappa e^{-\kappa t}-\delta(t)\right) & (t\ge0) \\
-\delta(t) & (t<0)
\end{cases}.
\end{align}
Thus, by convoluting $f_0(t)$ with $g(-t)$, and then setting the normalization constant properly, we obtain the optically-high-pass-filtered mode function as
\begin{equation}
f_{\text{HPF}}(t)=\sqrt{\gamma}\times
\begin{cases}
e^{-\gamma t} & (t\ge0) \\
\left(\frac{\gamma+\kappa}{\gamma-\kappa}e^{\gamma t}-\frac{2\kappa}{\gamma-\kappa}e^{\kappa t}\right) & (t<0)
\end{cases}.
\label{eq:mode_function}
\end{equation}
The temporal shapes of Eq.~(\ref{eq:mode_function}) are shown in Fig.~\ref{fig:mode_function}.
Contrary to the original non-negative mode function, this function has a negative dip due to its lack of low frequency components.

\section{Experimental setup}\label{sec:setup}

\begin{figure}[!tb]
  \begin{center}
    \includegraphics[width=\linewidth]{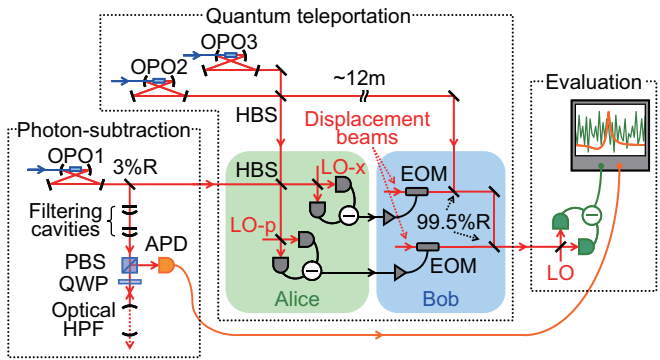}
  \end{center}
  \vspace{-4mm}
  \caption{(Color online) Experimental setup. APD: Avalanche photo-diode, EOM: Electro-optic modulator, HBS: Half beam splitter, LO: Local oscillator, OPO: Optical parametric oscillator, PBS: Polarizing beam splitter, QWP: Quarter wave plate.}
  \label{fig:experimental_setup}
\end{figure}

Our experimental setup consists of three parts; the first part is for photon-subtraction, the second part for quantum teleportation, and the third part for 
evaluating the quality of the teleported states (Fig.~\ref{fig:experimental_setup}).
Apart from the optical HPF part, the setup is mostly identical to that in Ref.~\cite{11Lee}, where more details are provided.

The output of a single-mode continuous-wave Ti:sapphire laser (SolsTiS-SRX, M Squared Lasers)
around 860 nm is used as the light source, a part of which is frequency-doubled for pumping three optical parametric oscillators (OPO). 
First, in the photon-subtraction part, a weakly squeezed vacuum is generated from OPO1 (half width at half maximum: 6.2 MHz) with a 15 mW pump.
The output of OPO1 is directed into a TBS which reflects 3\% of the beam toward an APD as the trigger beam, while the other beam is sent to the teleportation setup as the signal beam.
In the trigger channel, photons in all of the modes except for the degenerate squeezed mode are blocked with two Fabry-Perot filtering cavities with much wider HWHM (55 and 18 MHz) than that of OPO1.
Then the trigger beam is injected into an optical HPF, which is a Fabry-Perot cavity with a round-trip length of 280 mm (HWHM: 500 kHz, Finesse: 1080).
Here the theoretical cutoff $\kappa$ is set to $2\pi\times0.5$ MHz.
By increasing $\kappa$, we can reduce the low-frequency noise contamination.
However, this decreases the event-to-dark-count ratio on the APD, which degrades the purity of the generated states.
Therefore the choice of $\kappa$ is a compromise between these two effects. 
The reflection of the optical HPF is mode-reshaped and sent to the APD, while the transmission is monitored on a photodiode for locking the cavity.
Ideally, the cavity perfectly blocks the $\omega_0$ component when locked on resonance, but in our setup 14\% of the reflected component remains due to a slight mismatch between the transmittances of two cavity mirrors (impedance mismatch).
This effect can be taken into account by a more detailed model of the mode function, which is described in the Appendix.
Conditioned on an APD click, a photon-subtracted squeezed vacuum is generated in the signal beam and used as an input state for the subsequent quantum teleportation circuit.

For realizing quantum teleportation of this state, first Alice and Bob need to share a pair of broadband EPR-correlated beams as a resource.
This pair can be obtained by mixing two strongly squeezed vacua from OPO2 and OPO3 (HWHM: 12 MHz) on a half beam splitter.
Alice combines the input state with one of the EPR pairs on a half beam splitter, then measures the $x$ and $p$ quadratures of the resulting two outputs.
The measurement results are sent to Bob as gain-tuned electric signals, which Bob uses to modulate displacement beams (shown in Fig.~\ref{fig:experimental_setup}) and to correct the other beam of the EPR pair. As a result, the input state is reconstructed.
In order to faithfully teleport the quantum state within the original wave packet quantum mode, all these operations are performed over a frequency spectrum broader than this wave packet spectrum.

In order to evaluate the quality of the teleported quantum states, the output beam is measured by a balanced homodyne detector.
The local oscillator phase in this detector is scanned for the full angular resolved
tomography of the state, and 100,000 quadrature data are recorded over $[0, 2\pi]$.
We use the inverse Radon transform based on polynomial series expansion~\cite{11Benichi-2} to reconstruct the Wigner functions $W(x,p)$.

In this teleportation circuit, and in other teleportation-based circuits as well, low-frequencies can be contaminated by mainly two sources of noise.
The first source is the laser noise around the carrier frequency.
The non-Gaussian states can be deteriorated by this noise derived from the local oscillator beams in imperfectly-balanced homodyne detections for Alice and the final evaluation, as well as the displacement beams directly mixed by Bob.
The second source is the low-frequency electric noise, which contaminates Alice's homodyne signal for the displacement operation and the final homodyne signal for the evaluation.
These types of noise often need to be removed with electrical HPFs.
In our setup, a HPF with a cutoff frequency of 101 kHz is required in the final homodyne signal.
Non-Gaussian states generated \textit{without} mode-filtering have low-frequency components, which are inevitably removed by this HPF. 
The resulting frequency mode mismatch in the homodyne-detection directly deteriorates the non-Gaussian states, thereby limiting the quality of the teleportation.
In contrast, non-Gaussian states generated \textit{with} mode-filtering do not have frequency components lower than 101 kHz, thus we can avoid the low-frequency noise contamination, as well as achieve almost perfect mode matching in the final homodyne-detection.


\section{non-Gaussian state generation with mode-filtering}\label{sec:photon-subtraction}

First the states generated by photon-subtraction are analyzed to demonstrate the mode-filtering effect by the optical HPF.
We record 200,000 frames of quadrature data both in the conventional and mode-filtering photon-subtraction.
When using the conventional method, we unlock the optical HPF and monitor the signal of the transmitted beam to keep it totally off resonant during the measurement time.

\begin{figure*}[!tb]
\centering
\subfigure[A Wigner function with the conventional method. $\gamma/2\pi=6.5$ MHz, $\kappa/2\pi=0$ MHz, $W(0,0)=-0.171\pm0.003$.]{
\hspace{1mm}\includegraphics[clip,scale=0.39]{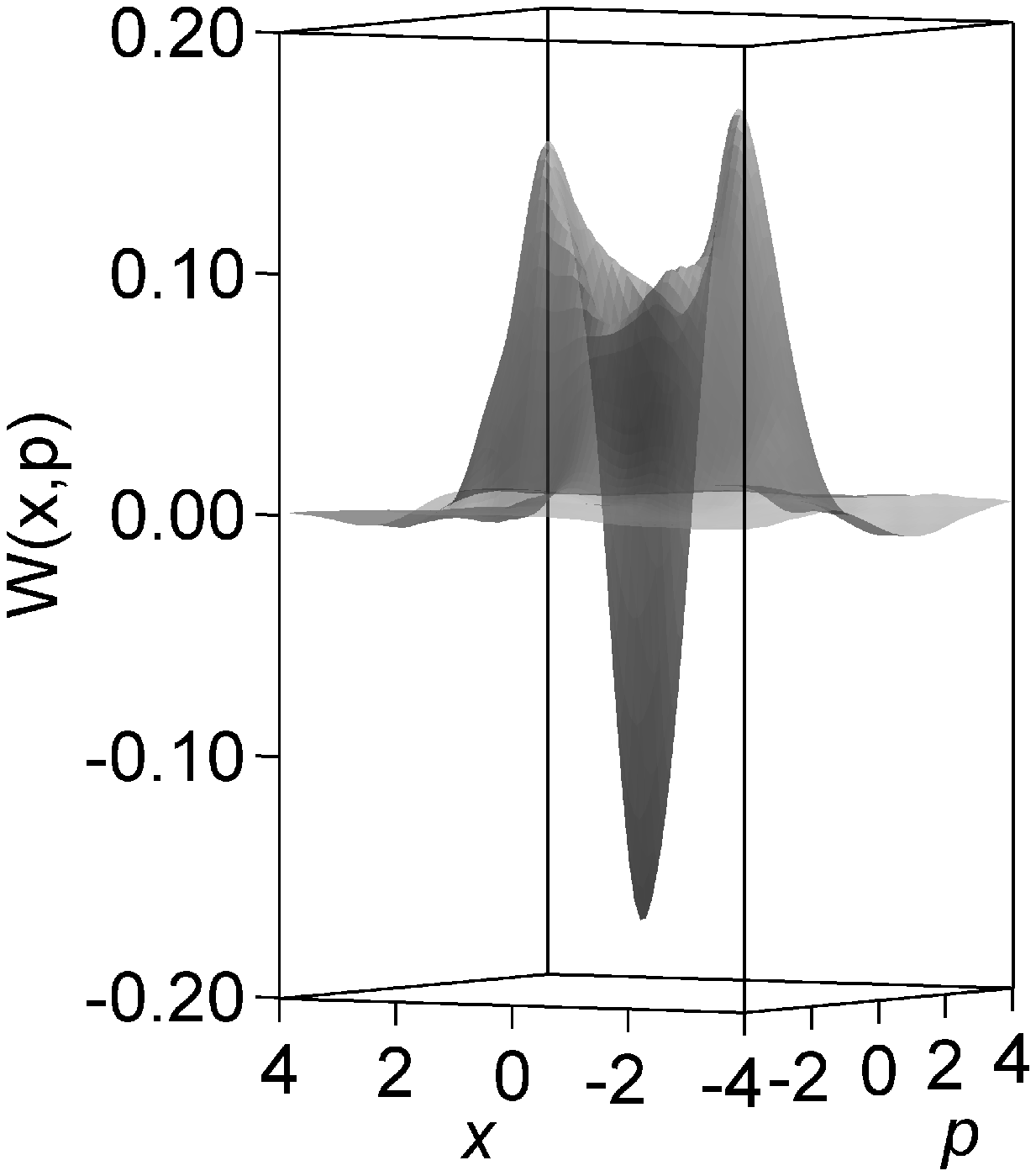}\hspace{1mm}
\label{sfig:input_full}
}
\subfigure[A Wigner function with the mode-filtering method. $\gamma/2\pi=6.5$ MHz, $\kappa/2\pi=0.48$ MHz, $W(0,0)=-0.179\pm0.003$.]{
\hspace{1mm}\includegraphics[clip,scale=0.39]{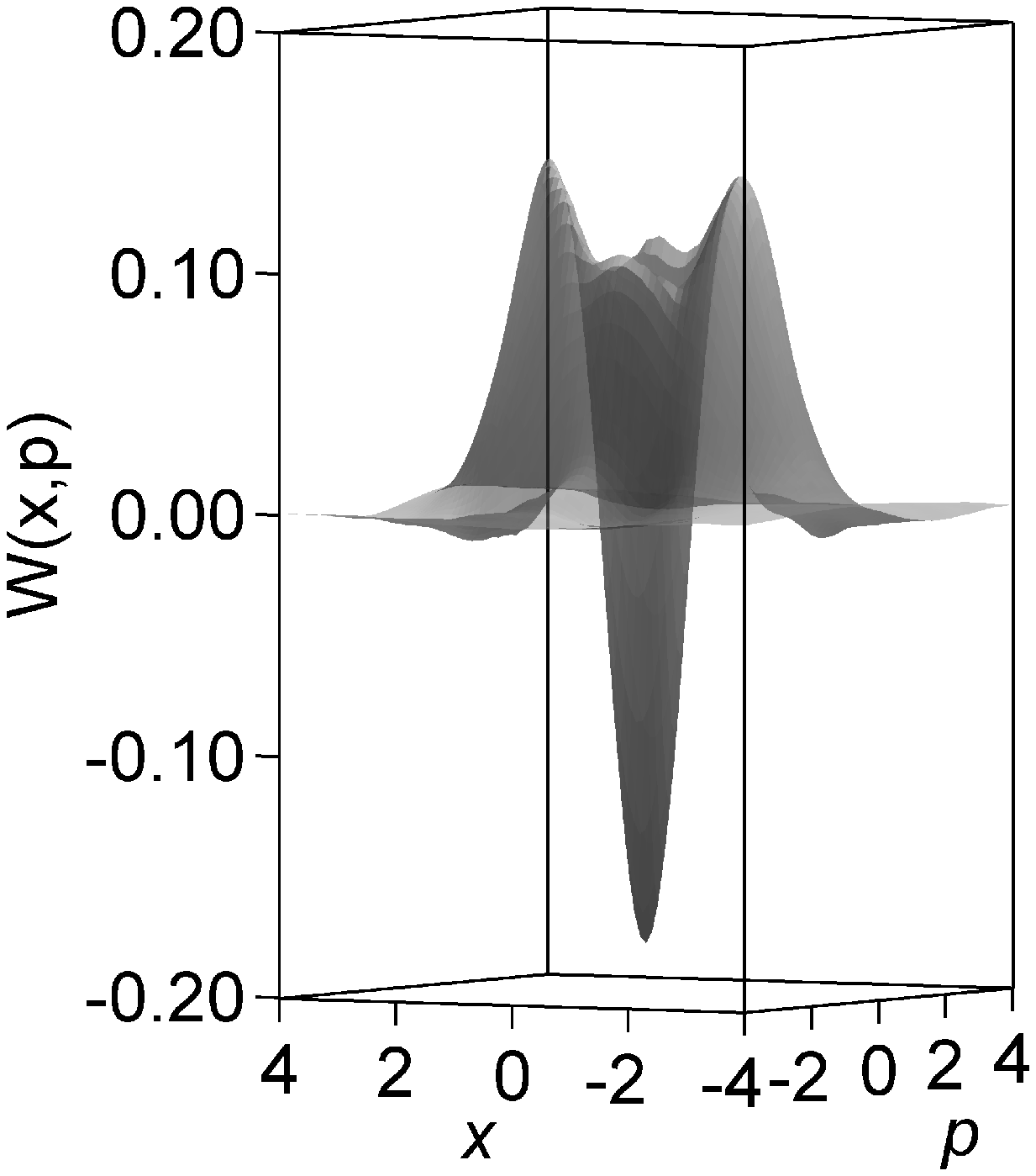}\hspace{1mm}
\label{sfig:input_ohp}
}
\subfigure[The dependence of $W(0,0)$ on the cutoff $\kappa$ of the applied mode function ($\gamma/2\pi=6.5$ MHz). Theoretical curves based on Eq.~(\ref{eq:negativity_theory}) are plotted as well.]{
\hspace{1mm}\includegraphics[clip,scale=1.3]{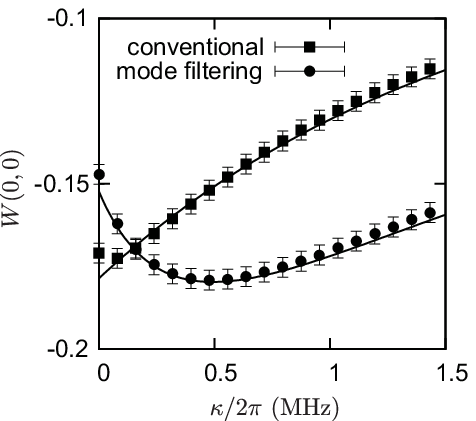}\hspace{1mm}
\label{sfig:input_negativity}
}
\caption{Experimental results of photon-subtraction with or without mode-filtering.}
\label{fig:input_results}
\end{figure*}

Experimental Wigner functions $W(x,p)$ in these two methods are shown in Fig.~\ref{sfig:input_full} and \ref{sfig:input_ohp}.
Here we applied the mode function including the effect of the impedance mismatch of
the optical HPF (Eq.~(\ref{eq:mode_function_mismatch}) in the Appendix).
Each Wigner function is reconstructed by optimizing the parameters $\gamma$ and $\kappa$ of the mode function to extract the minimum $W(0,0)$ ($\kappa$ is set to 0 for the conventional method). 
The optimal parameters of $\gamma=2\pi\times6.5$ MHz and $\kappa=2\pi\times0.48$ MHz match well the experimentally estimated values of $2\pi\times6.2$ MHz (OPO1 bandwidth) and $2\pi\times0.50$ MHz (optical HPF bandwidth).
Both Wigner functions have minimum negativities of $W(0,0)=-0.171\pm0.003$ (conventional) and $W(0,0)=-0.179\pm0.003$ (mode-filtering) without any corrections,
thereby excluding any description as a classical probability density. 
The experimental Wigner functions sometimes have minimum values not exactly at the origin, but at a slightly shifted point around the origin
possibly due to the interference of the trigger photons with the other unwanted fake trigger photons.
Here, we adopt the minimum negativity around the origin as $W(0,0)$.
This non-classical negative region is fragile and easily disappears with every percent of experimental loss and other imperfections.
Thus, the quality of these non-Gaussian states can be evaluated by their negativity.

In Fig.~\ref{sfig:input_negativity}, we plot the dependence of the minimum value $W(0,0)$ on the cutoff $\kappa$ of applied mode functions ($\gamma$ is fixed at $\gamma=2\pi\times6.5$ MHz).
Without the optical HPF, the theoretical mode function $f_0(t)=\sqrt{\gamma}e^{-\gamma|t|}$ corresponds to the case of $\kappa=0$ in Eq.~(\ref{eq:mode_function_mismatch}).
Therefore $W(0,0)$ increases with increasing $\kappa$ as a result of the mode mismatch between the photon-subtracted temporal mode and the applied mode function.
With the optical HPF, the plot shows the different $\kappa$ dependence which has an optimal $\kappa$.

These experimental data are in good agreement with the theoretical curves calculated from a simple and realistic model~\cite{11Benichi-1}, thereby showing the experimental success of the mode-filtering (Fig.~\ref{sfig:input_negativity}).
In the theoretical model, the generated state is modeled by $\hat{a}\hat{S}(r)\ket{0}$, where $\hat{a}$ is the annihilation operator describing photon subtraction and $\hat{S}(r)$ is the squeezing operator with a squeezing parameter $r$.
This state is degraded by experimental losses and imperfections, and as a result the value of the Wigner function at the origin can be expressed as~\cite{11Benichi-1}
\begin{equation}
W(0,0)=\frac{1-2\eta+2\zeta\eta\left[1+2(1-\eta)\sinh^2(r)\right]}
{\pi\left[1+4\eta(1-\eta)\sinh^2(r)\right]^{3/2}},
\label{eq:negativity_theory}
\end{equation}
where $\eta$ is the overall efficiency and $\zeta$ is given as a dark count rate divided by a total count rate.
The ideal limit of $\eta=1$ and $\zeta=0$ gives $W(0,0)=-1/\pi$ regardless of $r$.

From our analysis the squeezing parameters were estimated as $r=0.38\pm0.02$ (conventional) and $r=0.36\pm0.02$ (mode-filtering).
The total count rates were $8200 \pm 300$ and $7200 \pm 300$ per second respectively, in which the dark count rate of $150\pm30$ per second was included.
Therefore $\zeta=0.02$ for each case.
The overall efficiency can be calculated as $\eta=\eta_{0}(\eta_\kappa)^2$, where $\eta_0$ is the efficiency before applying a mode function, and $\eta_{\kappa}=\int f_\text{opt}(t)f_\kappa(t)dt$ is a mode matching parameter between the optimal mode function $f_\text{opt}(t)$ and the applied mode function $f_\kappa(t)$.
Using the parameters above, we calculated $W(0,0)$ for each $\kappa$ by fixing the unknown parameter $\eta_0$ for the most plausible value of $0.83$.

$\eta_0$ was also calculated directly from experimentally estimated values as
$\eta_0=\eta_{\text{OPO}}\eta_{\text{pr}}(\eta_{\text{vis}})^2\eta_{\text{hom}}$,
where the OPO escape efficiency $\eta_{\text{OPO}}=0.98$, the propagation efficiency $\eta_{\text{pr}}=0.95$, the visibility $\eta_{\text{vis}}=0.98$, the homodyne efficiency $\eta_{\text{hom}}=0.95$ which includes the efficiency of the photodiode and the electrical signal-to-noise ratio.
All these values gave $\eta_0=0.85$, which is reasonably close to the directly estimated value of $0.83$.

It appears there is some discrepancy between experiment and theory at $\kappa=0$ in Fig.~\ref{sfig:input_negativity},
which we partly ascribe to some low frequency noise.
It implies that the negativity in the conventional method is partly limited by this noise, whereas in the mode-filtering method the mode does not have low frequency components and thus can avoid this limitation.

These data and analysis show the successful mode-filtering as predicted from the theory.
Therefore, we succeeded in tailoring the quantum mode of non-Gaussian states, while preserving or even enhancing the non-classicality of the state within the mode.


\section{Quantum teleportation with mode-filtering}\label{sec:teleportation}

\begin{figure*}[!tb]
\centering
\subfigure[$W(0,0)$ of four data sets. Averaged values are plotted with dashed (mode filtering) and dot-dashed (conventional) lines.]{
\hspace{1mm}\includegraphics[clip,scale=1.15]{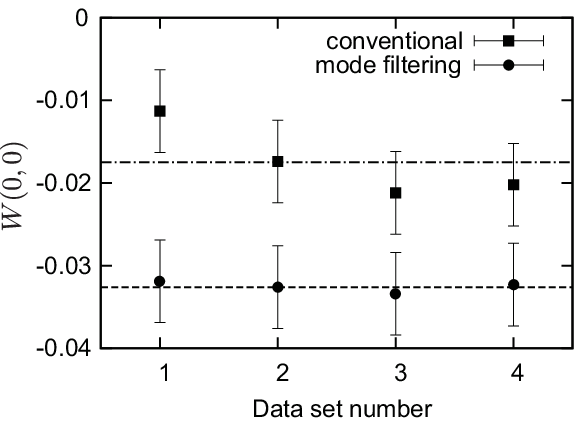}\hspace{1mm}
\label{sfig:output_negativity}
}
\subfigure[A Wigner function with the conventional method. $W(0,0)=-0.017\pm0.005$.]{
\hspace{1mm}\includegraphics[clip,scale=0.39]{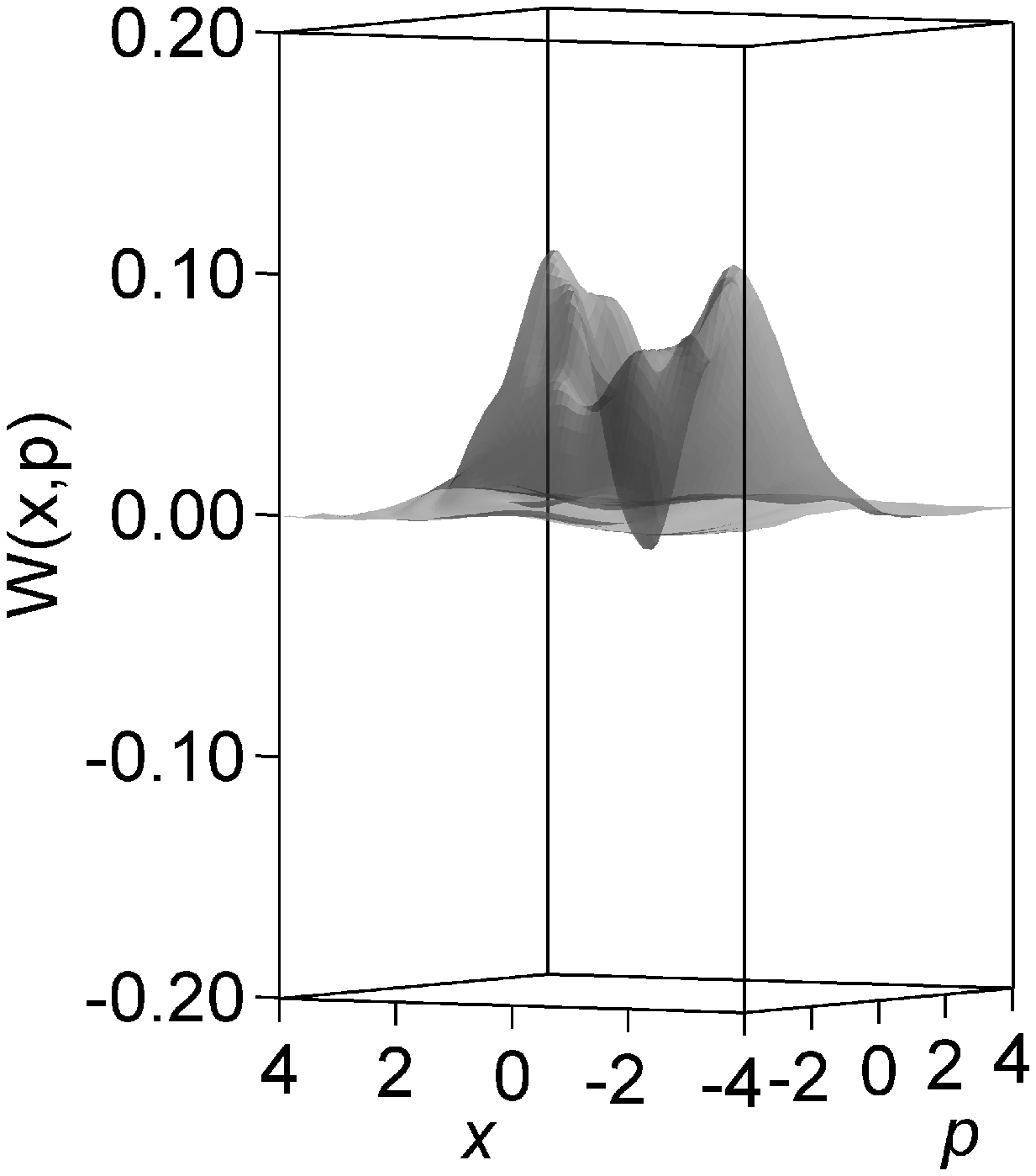}\hspace{1mm}
\label{sfig:output_full}
}
\subfigure[A Wigner function with the mode-filtering method. $W(0,0)=-0.033\pm0.005$.]{
\hspace{1mm}\includegraphics[clip,scale=0.39]{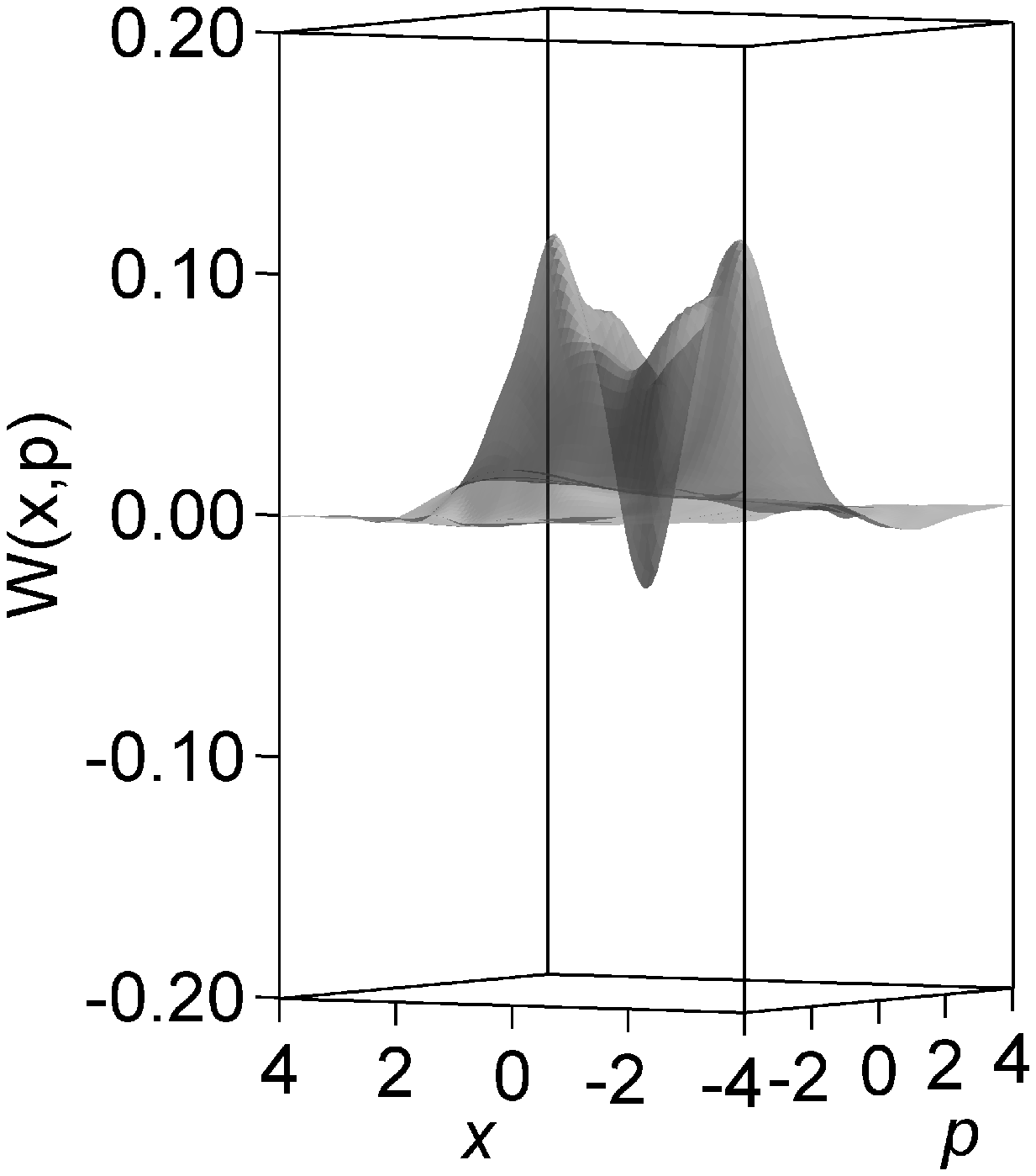}\hspace{1mm}
\label{sfig:output_ohp}
}
\caption{Experimental results of quantum teleportation with or without mode-filtering. (b) and (c) correspond to ``Data set number 2'' of (a).}
\label{fig:output_results}
\end{figure*}

The next step is to demonstrate the advantage of our mode-filtering method by measuring the degree of non-classicality preserved in the state after teleportation.
For this purpose a tomography data set of the teleported states with and without the mode-filtering (100,000 frames each) is measured 4 times in a row.
We reconstructed Wigner functions from each data set with optimized parameters of mode functions.
The minimum negativities $W(0,0)$ of all data sets are shown in Fig.~\ref{sfig:output_negativity}, and a selection of the teleported Wigner functions are shown in Fig.~\ref{sfig:output_full} and \ref{sfig:output_ohp}.

All of the Wigner functions exhibit negativities around the origin, showing the successful teleportation of the fragile non-classical feature.
Negative values are estimated from four sets of reconstructed Winger functions as $W(0,0)=-0.018\pm0.007$ (conventional) and $W(0,0)=-0.033\pm0.005$ (mode-filtering).
The Wigner functions in Fig.~\ref{sfig:output_full} and \ref{sfig:output_ohp} exhibit a clear difference in the depth at the center dip.
The mode-filtered states explicitly preserve stronger negativities, demonstrating a clear filtering-related enhancement of the states' robustness in the teleportation process.
Due to the impedance mismatch of the optical HPF, our mode-filtered states still have some low-frequency components, and are therefore partly affected by the noise contamination.
If perfect impedance matching is obtained, the enhancement can be even greater.

In summary, we have experimentally demonstrated that our efficient encoding strategy with mode-filtering is highly effective in the current teleportation apparatus, where low frequencies are contaminated by the laser noise and electric noise in three homodyne measurements and two displacement operations.
This filtering-related enhancement can also be seen in a wide range of CV protocols, especially teleportation-based protocols with the same kind of inevitable noise contamination in practice.


\section{Conclusion}\label{sec:conclusion}

In conclusion, we have developed and demonstrated a mode-filtering technique of non-Gaussian states generated by photon-subtraction.
By introducing a reflecting cavity in the trigger channel, we removed the problematic noisy low-frequency components of the non-Gaussian states.
The mode-reshaped states are analyzed with the mode function derived from a theoretical model, and the agreement of experiment and theory shows the successful mode-filtering without any degradation of the non-classical states within the mode.
After the teleportation, the mode-filtered states show stronger non-classicality than the states without mode-filtering, thereby showing that our technique is effective to achieve robust implementation of CV protocols, especially teleportation-based protocols utilizing non-Gaussian states. 
This technique can be straightforwardly applied to various types of photon-subtraction protocols, and thus it should be an important ingredient in a variety of applications of non-Gaussian states towards universal CV QIP.


\section*{APPENDIX: THE EFFECT OF IMPEDANCE MISMATCH}\label{sec:corrected_mode_function}

In general, the reflected mode of the optical HPF can be written in the frequency domain as~\cite{00Lu}
\begin{align}
\hat{a}_{\text{r}}(\omega_0+\omega)=&\frac{(\kappa_1-\kappa_2)/2+i\omega}{(\kappa_1+\kappa_2)/2-i\omega}\hat{a}_{\text{in}}(\omega_0+\omega)\nonumber\\
&+\frac{\sqrt{\kappa_1\kappa_2}}{(\kappa_1+\kappa_2)/2-i\omega}\hat{a}_{\text{v}}(\omega_0+\omega),
\label{eq:cavity_relation_k1k2}
\end{align}
where $\kappa_1$ and $\kappa_2$ denote the decay rates of two cavity mirrors.
If $\kappa_1=\kappa_2=\kappa$, the impedance match is perfect and Eq.~(\ref{eq:cavity_relation_k1k2}) gives Eq.~(\ref{eq:cavity_relation}).
Otherwise we define $\kappa\equiv(\kappa_1+\kappa_2)/2$ and $\kappa^\prime\equiv(\kappa_1-\kappa_2)/2$, then the response function of the cavity is given by $\tilde{g}(\omega)=(\kappa^\prime+i\omega)/(\kappa-i\omega)$. Since $\tilde{g}(0)\not=0$, some part of the $\omega_0$ component is reflected toward the APD.
By convoluting $\tilde{g}(-t)$ with the original mode function $f_0(t)=\sqrt{\gamma}e^{-\gamma|t|}$, we obtain
\begin{equation}
\tilde{f}_{\text{HPF}}(t)=N\times
\begin{cases}
\frac{\gamma-\kappa^\prime}{\gamma+\kappa}e^{-\gamma t} & (t\ge0) \\
\left(\frac{\gamma+\kappa^\prime}{\gamma-\kappa}e^{\gamma t}-\frac{2\gamma(\kappa+\kappa^\prime)}{\gamma^2-\kappa^2}e^{\kappa t}\right) & (t<0)
\end{cases},\label{eq:mode_function_mismatch}
\end{equation}
where $N$ is a normalization constant given by
\begin{equation}
N=\sqrt{\frac{\gamma\kappa(\gamma+\kappa)^2}{\gamma^2\kappa+\kappa^{\prime 2}(2\gamma+\kappa)}}.
\end{equation}
In this experiment, $\kappa^\prime$ is set to $\kappa^\prime=-0.37\kappa$ based on experimental values.

\section*{ACKNOWLEDGEMENTS}

This work was partly supported by the SCOPE program of the MIC of Japan, PDIS, GIA, G-COE, APSA, and FIRST commissioned by the MEXT of Japan, and ASCR-JSPS.
The authors acknowledge Seiji Armstrong for his assistance with the editing of this paper.


\end{document}